\documentclass{aa}  

\usepackage[colorlinks=true,
		    linkcolor=blue,
		    citecolor=blue]{hyperref}
\usepackage{graphicx}
\usepackage{txfonts}
\usepackage{lipsum}
\usepackage{subcaption}         
\usepackage{lscape}             
\usepackage{placeins}           
\usepackage{natbib}
\bibpunct{(}{)}{;}{a}{}{,} 

\usepackage[dvipsnames]{xcolor}
\usepackage{amssymb}
\usepackage{bbm}
\usepackage{amsmath}
\usepackage{mathtools}
\usepackage{fontawesome}
\usepackage{hhline}
\usepackage{multirow}
\usepackage{dblfloatfix}
\usepackage{yhmath}

\newcommand{\phir}{\phi_\text{r}}
\newcommand{\dphi}{\delta\phi}
\newcommand{\dphir}{\dphi\!+\!\phir}

\DeclareFontFamily{U}{mathx}{}
\DeclareFontShape{U}{mathx}{m}{n}{<-> mathx10}{}
\DeclareSymbolFont{mathx}{U}{mathx}{m}{n}
\DeclareMathAccent{\widehat}{0}{mathx}{"70}
\DeclareMathAccent{\widecheck}{0}{mathx}{"71}
\DeclareSymbolFont{yhlargesymbols}{OMX}{yhex}{m}{n}
\DeclareMathAccent{\yhwidehat}{\mathord}{yhlargesymbols}{"62}
\renewcommand{\widehat}{\yhwidehat}

\newcommand{\pscal}[2]{\langle \; {#1} \, | \, {#2} \; \rangle}

\definecolor{Pink}{HTML}{E75874}

\definecolor{greencheck}{HTML}{2B7C0D}
\definecolor{reduncheck}{HTML}{A73118}
\newcommand{\checksymb}{\textcolor{greencheck}{\faCheckCircle}}
\newcommand{\unchecksymb}{\textcolor{reduncheck}{\faTimesCircle}}

\begin{document}

   \title{Optical gains measurement with a gain scheduling camera: On-sky demonstration with PAPYRUS and perspectives}


   \author{A. Striffling\inst{1}\fnmsep\inst{2}\fnmsep\inst{3} \thanks{Corresponding author, \email{arnaud.striffling@lam.fr}} \and
   		  C.-T. Héritier\inst{1}\fnmsep\inst{3} \and
   		  R.J.-L Fétick\inst{1}\fnmsep\inst{3} \and
   		  O. Fauvarque\inst{4} \and
   		  J.-F Sauvage\inst{1}\fnmsep\inst{3} \and
   		  A. Carlotti\inst{2} \and
   		  B. Neichel\inst{3} \and
   		  T. Fusco\inst{5}\fnmsep\inst{3}
          }

   \institute{DOTA, ONERA, 13330, Salon-de-Provence, France \and
                Univ. Grenoble Alpes, CNRS, IPAG, 38000 Grenoble, France \and
   			Aix Marseille University, CNRS, CNES, LAM, Marseille, France \and
   			Ifremer, RDT Research and Technological Development, Plouzané, France \and
   			DOTA, ONERA, Université Paris Saclay, 91120, Palaiseau, France}

    \date{}
    
  \abstract
   {Reaching the high angular resolution and contrast level desired for exoplanetary science requires us to equip large telescopes with extreme adaptive optics (XAO) systems to compensate for the effect of the atmospheric turbulence at a very fast rate. This calls for the development of ultra-sensitive wavefront sensors (WFSs), such as Fourier filtering wavefront sensors (FFWFSs), to be operated at low flux, as well as an increase in the XAO loop frame rate. These sensors, which constitute the baseline for current and future XAO systems, exhibit such a high sensitivity at the expense of a non-linear behaviour that must be properly calibrated and compensated for to deliver the required performance.}
 {We aim to validate on-sky a recently proposed method that associates the FFWFS with a focal plane detector -- the gain scheduling camera (GSC) -- to estimate in real time the first-order terms of the sensor non-linearities, known as modal optical gains.}
   {We implemented a GSC on the adaptive-optics (AO) bench PAPYRUS to be associated with the existing pyramid wavefront sensor (PWFS). We compared experimental results to expected results obtained with a high-fidelity numerical twin of the AO system.}
   {We validated experimentally the method both in laboratory and on-sky. We demonstrated the capability of the GSC to accurately estimate the optical gains of the PWFS at 100~Hz, corresponding to the current limit in speed imposed by PAPYRUS hardware, but it could be applied at higher frequencies to enable frame-by-frame optical gains tracking. The presented results exhibit good agreement on the optical gains estimation with respect to numerical simulations reproducing the experimental conditions tested.}
   {Our experimental results validate the strategy of coupling a FFWFS with a focal-plane camera to master the non-linearities of the sensor. This demonstrates its attractiveness for future XAO application.}

               
    \authorrunning{A. Striffling et al.}
    \titlerunning{On-sky optical gains measurement with the gain scheduling camera}
    
   \maketitle

  \section{Introduction}

The unprecedented angular resolution of the future Extremely Large Telescope (ELT) \citep{gilmozzi2007} will revolutionise the field of exoplanetary science. The development of medium- (1K–10K) to high-resolution (10K–100K) spectroscopic instruments \citep{thatte2024,brandl2024,marconi2024a} will enable exoplanets' compositions to be probed through direct imaging, with the aim of detecting biosignatures. Achieving this ambitious goal is restricted to planets orbiting nearby stars -- including non-transiting ones -- by spatially, and hence angularly, separating the light from the star and that from the planet. Angularly resolving an exoplanet from its host star presents two main instrumental challenges. First, it demands a very high angular resolution to disentangle the signals of the planet and of the host star, with separations ranging from tens to hundreds of milliarcseconds for rocky planets \citep{chauvin2024}. Second, it requires extremely high-contrast capabilities, as the flux ratio between star and planet ranges from $10^{-6}$ for young Jupiters to $10^{-9}$ for super-Earths \citep{kasper2021}.

Although the angular resolution should be granted by the colossal diameter of the future ELT, telescopes are affected by the phase aberrations induced along the light propagation by the atmosphere, the telescope, and the imaging instrument's optics. These aberrations are responsible for a dramatic loss of angular resolution and of signal-to-noise ratio compared to a diffraction-limited system. An instrument is considered diffraction-limited if the wavefront error is below $\lambda/14$ \citep{born1999}. Although this criterion suffices to restore maximum spatial resolution, further improvements are necessary to meet the demands of high-contrast imaging, since even modest increases in the Strehl ratio (SR) critically enhances coronagraphic performance. Specifically, the intensity of residual stellar speckles -- and thus stray light -- scales approximately with the square of the wavefront error \citep{fusco2006}. Therefore, pushing adaptive optics (AO) systems towards extreme wavefront correction is essential.

Thanks to a powerful extreme AO (XAO) system \citep{fusco2006,beuzit2019}, the SPHERE instrument has successfully imaged giant (Jupiter-like) exoplanets located far from their host stars. To achieve this, the XAO system reduced the residual wavefront error to a few tens of nanometres root mean square (RMS) and, in combination with coronagraphy, reached contrast levels of approximately $10^{-6}$. Although impressive, such contrast remains insufficient for the detection of Earth-like planets. An ongoing upgrade of SPHERE \citep{boccaletti2022} is under development to implement a second-stage AO system designed to improve contrast by tackling the dominant source of wavefront error: the temporal error. Similarly, the Gemini Planet Imager (GPI), which achieves a comparable contrast performance, will soon be succeeded by GPI 2.0, targeting contrast levels of up to $10^{-7}$ \citep{chilcote2024}. However, accessing exoplanets that may exhibit biosignatures will require the ELT’s angular resolution, combined with a very fast XAO system such as that planned for the Planetary Camera and Spectrograph (PCS) instrument \citep{kasper2021}.

To mitigate the temporal error, XAO instruments must operate at very high speeds, which results in lower signal-to-noise ratios and necessitates the use of highly sensitive wavefront sensors (WFSs). In this context, several studies have demonstrated the significant sensitivity gain offered by Fourier-filtering wavefront sensors (FFWFSs) compared to the Shack–Hartmann sensor \citep{verinaud2004,plantet2015}. Among these, the pyramid wavefront sensor (PWFS) is of particular interest, though it exhibits strong non-linear behaviour that must be addressed to achieve optimal reconstruction of the phase in operation, to perform off-zero operations as described in Sect.~\ref{sec:off-zero_cases} or to retrieve the point spread function (PSF) with the AO telemetry \citep{veran1997a}. Following the same logic as other concepts making use of multiple optical planes, such as the non-linear curvature WFS \citep{guyon2010}, an enhanced version of the PWFS has been proposed: the focal-plane assisted PWFS \citep{chambouleyron2021}. This approach integrates an additional focal-plane measurement to estimate a first-order approximation of the non-linear effects, referred to as optical gains \citep{korkiakoski2008a}. Initially developed through theoretical analysis and validated in numerical simulation, this concept has yet to be experimentally demonstrated in either laboratory or on-sky conditions. 

The present study details the integration of the focal-plane assisted PWFS, covering aspects from optical hardware implementation to on-sky optical gain measurements, alongside the associated algorithmic developments. Validation is conducted both in the laboratory and on-sky using the PAPYRUS instrument \citep{fetick2023}, an AO test bed installed on the 1.52~m telescope at the Observatoire de Haute-Provence (OHP). Section~\ref{sec:pwfs} reviews the various definitions of optical gains proposed in the literature, the current estimation methods, and the advantages of operating the PWFS away from its zero set point. Section~\ref{sec:FPAPWS} describes the implementation of a gain scheduling camera on PAPYRUS, together with the experimental optical gains measurements obtained under controlled conditions using the internal calibration source. Finally, Sect.~\ref{sec:onsky} presents the first on-sky optical gains measurements acquired via the gain scheduling camera for different observing conditions.

\section{The pyramid wavefront sensor: A non-linear sensor}
\label{sec:pwfs}

The PWFS is one of the FFWFSs \citep{fauvarque2016}. The mask is a pyramid-shaped optical component made of glass with its tip located in the focal-plane. Each tilted quadrant of the pyramid angularly separates one quarter of the incoming beam. The pyramid thus divides the incident beam into four sub-beams, re-imaged on a detector placed in pupil plane. The optical layout is presented in Fig.~\ref{fig:fppwfs}. The number of pupils imaged and their separation depend, respectively, on the number of faces and the angle of the pyramid. Different configurations have been studied by the community to  mitigate the sensitivity and linearity \citep{fauvarque2015a,janin-potiron2019,schatz2021,chambouleyron2023b}, but the four-sided pyramid, introduced by \citet{ragazzoni1996}, remains the most common implementation so far. As such, we focus in this paper on the original implementation with four faces even though the formalism applies to any FFWFS.

The PWFS acts on the electromagnetic field with the pyramid-shaped optical component and captures a useful signal, $I(\phi)$, thanks to its pupil plane detector. We usually define a signal of interest, $\Delta I(\dphi{;}\phir)$, as the difference between the raw intensity signal of the perturbation, $I(\phi)$, with $\phi=\dphir$, and a reference intensity signal obtained around a reference phase, $\phir$, $I(\phir)$, after a normalisation based on the flux, such as
\begin{equation}
    \Delta I(\dphi{;}\phir) = \frac{I(\phi)}{N_\text{ph}} - \frac{I(\phir)}{N_\text{ph}} = \frac{I(\dphir)}{N_\text{ph}} - \frac{I(\phir)}{N_\text{ph}}
,\end{equation}
where we consider an invariant flux of $N_\text{ph}$ photons per frame. Based on the pixel illumination, a selection of the pixels of interest is achieved to extract the useful signal from the detector. It is then possible to keep this reduced signal as it is: the pyramid is working in intensity maps, or to reduce it further to work in slopes maps \citep{verinaud2004}.  In this paper we consider the intensity maps, as they represent the most straightforward approach to process PWFS signal. However, this choice is system-dependent and does not affect the optical gains estimation.

As any type of FFWFS, the PWFS exhibits a non-linear regime when the input phase is too large with respect to the sensing wavelength, $\Delta I(\dphi;\phir)$ no longer evolves proportionally to $\dphi$. The linearity range of a PWFS can be increased by applying a tip-tilt modulation at the cost of a lowered sensitivity \citep{esposito2001,verinaud2004}. In the linear regime, the reconstruction of the phase can be computed making use of the calibration of the PWFS. This calibration is usually achieved by measuring the interaction matrix, \textbf{M}, of the system that is then inverted to provide the reconstruction matrix. The measurement of the interaction matrix is commonly obtained using the so-called push-pull method, mirroring the finite-difference derivative method, to provide the local derivative, $\delta I$, of $I$ with respect to a phase mode, $Z_i$, around the working point, $\phir$:
\begin{equation}
    \delta I(\epsilon Z_i{;}\phir)\ = \frac{\Delta I(\epsilon Z_i{;} \phir) - \Delta I(-\epsilon Z_i{;} \phir)}{2\epsilon}
    \label{eq:delta_i_pushpull}
,\end{equation}
where $\epsilon$ is the input phase mode amplitude, chosen to be small enough to stay in the linear regime around $\phir$. This operation is repeated for all the modes, $Z_i$, of the control modal basis, $\{Z_i\}_m$. The choice of the modal basis (Zernike modes, Karhunen-Loève or deformable mirror influence functions) is usually driven by system constraints. The concatenation of the $\delta I$ gives the interaction matrix, \textbf{M}, measured around $\phir$:
\begin{equation}
	\textbf{M}(\phir)=\left(\vphantom{n^2}\delta I(Z_1{;}\phir),\ \hdots,\ \delta I(Z_N{;}\phir) \right) 
.\end{equation}
The matrix-based approach assumes a linear relationship between the incoming wavefront and the intensity pattern delivered by the PWFS:
\begin{align}
    \text{Propagation $\dphi\to I$}: & \quad \Delta I(\dphi{;}\phir) = \textbf{M}(\phir)\,\dphi \label{eq:linear_relationship_phase_intensity}\\
    \text{Reconstruction $I\to\dphi$}: & \quad \dphi = \textbf{M}^{+}\!(\phir)\,\Delta I(\dphi{;}\phir) \notag
,\end{align}
with $\cdot^{+}$ the pseudo-inverse. The validity of this assumption is thus only guaranteed in the linear regime, usually for low to moderate aberrations. Larger amplitudes entail a non-linear behaviour.

Non-linearities take different forms on the phase reconstruction. At first order, the impact is a modal-dependent loss of sensitivity of the PWFS -- known as the optical gains -- deteriorating the AO performance by underestimating the aberrations. Then, cross-talk between modes appears -- the modal confusion \citep{cisse2024} -- impairing the phase reconstruction, and thus the stability of the AO. In this work, we target a fast method, suitable for real-time computation during AO closed-loop, to estimate the main contribution of the non-linearities, and therefore only focus on the estimation of the modal optical gains. Such a method would enable the operation of the PWFS around a non-zero phase, for reasons described in the following section, and to target XAO performance.

\subsection{Offsetting the PWFS}
\label{sec:off-zero_cases}

The error budget of an XAO system is critical, to ensure the highest performance possible. The main terms such as the fitting and temporal errors are therefore minimised. Some errors, usually less limiting, become dominant. It is the case of the non-common path aberrations (NCPAs), arising from optical path differences between the WFS and the scientific branches \citep{sauvage2007a,vigan2022}. Different techniques already exist to compensate NCPA, such as the use of a phase plate \citep{clenet2019} or a dedicated secondary deformable mirror \citep{dohlen2018}. This compensation may also be achieved by offsetting the PWFS \citep{esposito2020,chambouleyron2023a}, which no longer works around a flat wavefront but around the NCPAs compensating phase. The proper introduction of NCPAs depends on the non-linearities of the PWFS, and thus on the precise and fast knowledge of the optical gains. Ignoring this effect alters the phase introduced and may ultimately lead to the so-called NCPAs catastrophe \citep{chambouleyron2021}; in the non-linear regime, the loss of sensitivity triggers an over introduction of the NCPAs compensating phase, the PWFS underestimating the offset truly applied. The excess of a created phase amplifies the effect of non-linearity, driving to a positive feedback causing the loop to diverge.

Operating the PWFS off-zero is also attractive to further optimise the scientific instrument capabilities with the introduction of controlled aberrations. Absolute control of the tip-tilt allows one to maximise the coupling of the PSF in a single-mode fibre, feeding a high-resolution spectrograph \citep{jovanovic2017,mawet2017,elmorsy2022a}. For direct imaging, the contrast can be enhanced by the introduction of dark hole map \citep{paul2014,bos2021,potier2022}, usually composed of higher-order modes.

\subsection{Defining the optical gains}
\label{sec:def_og}

Phase residuals spread the energy across the focal plane, in a manner analogous to tip-tilt modulation. Consequently, this effect can be interpreted as a form of self-modulation of the phase, which reduces the sensitivity of the PWFS. While tip-tilt modulation is well controlled and calibrated as part of the interaction matrix -- remaining stable during on-sky operation -- self-modulation is driven by the evolving wavefront residuals. This dynamic behaviour continuously modifies the PWFS sensitivity, making the interaction matrix unsuitable for accurate phase reconstruction. Although recalibrating around the residual phase could enhance accuracy, the required time makes this approach impractical during on-sky observations.

An alternative approach is proposed in \citet{chambouleyron2021}, which models the PWFS as a linear parameter-varying system (LPVS). In this framework, non-linearities modify the locally linear operating regime. Rather than recomputing the interaction matrix, it is retained and adjusted dynamically to the current non-linear state using transition matrices -- since $\textbf{M}(\phir)$ and $\textbf{M}(\dphir)$ have the same rank, and thus are equivalent -- such as
\begin{equation}
 \textbf{M}(\dphir) = \textbf{I}(\dphir)\,\textbf{M}(\phir)\,\textbf{G}(\dphir)
    \label{eq:M0_Mphi_link}
,\end{equation}
with \textbf{I}($\dphir$) and \textbf{G}($\dphir$) the transition matrices describing, respectively, the rearrangement in the space of intensities and phases. A first approximation is made; we call it $\mathcal{H}_1$, considering that, independently of the non-linearity strength, the signal on the PWFS detector can be interpreted as a linear combination of the responses, $\delta I$, of the interaction matrix, $\textbf{M}(\phir)$.
\begin{equation}
\mathcal{H}_1 \Rightarrow \textbf{M}(\dphir) = \textbf{M}(\phir)\,\textbf{G}(\dphir)
\end{equation}
And then, with the pseudo-inverse definition:
\begin{align}
\textbf{G}(\dphir) & = \textbf{M}^{+}(\phir)\,\textbf{M}(\dphir) \notag \\
    & = \left(\textbf{M}^\textsc{t}(\phir)\,\textbf{M}(\phir)\right)^{-1}\textbf{M}^\textsc{t}(\phir)\,\textbf{M}(\dphir)
    \label{eq:OG_matricial}
.\end{align}
The second approximation, $\mathcal{H}_2$, makes the assumption that the resultant of $\textbf{M}^\textsc{t}(\phir)\,\textbf{M}(\phir)$ is diagonal, and the third and last approximation, $\mathcal{H}_3$, is similar and assumes that $\textbf{M}^\textsc{t}(\phir)\,\textbf{M}(\dphir)$ is also diagonal. Those hypotheses imply complete orthogonality between the PWFS responses for different phase realisations, whatever the non-linearities. They remain valid in case of high diagonal ratio, where the cross-coupling between modes is negligible, which is typically the case under moderate non-linear behaviour such as that encountered in a closed-loop operation.
\begin{align}
\mathcal{H}_1, \mathcal{H}_2, \mathcal{H}_3 \Rightarrow \textbf{G}(\dphir) & = \frac{\text{diag}\left(\textbf{M}^\textsc{t}(\phir)\,\textbf{M}(\dphir)\right)}{\text{diag}\left(\vphantom{\phi}\textbf{M}^\textsc{t}(\phir)\,\textbf{M}(\phir)\right)}
    \label{eq:OG_matricial_hyp}
\end{align}
Accordingly, \textbf{G} is a diagonal matrix, with diagonal elements corresponding to the respective mode-to-mode optical gains. 

This definition slightly differs from the one originally introduced in \citet{korkiakoski2008a}, defined as
\begin{equation}
\textbf{G}_\textbf{K}(\dphir) = \sqrt{\frac{\text{diag}\left(\textbf{M}^\textsc{t}(\dphir)\ \textbf{M}(\dphir) \right)}{\text{diag}\left(\textbf{M}^\textsc{t}(\phir)\ \textbf{M}(\phir) \right)}}
,\end{equation}
where $\textbf{M}^\textsc{t}(\cdot)\textbf{M}(\cdot)$ are square matrices whose diagonal elements encode the squared sensitivity of each mode. This formulation expresses a ratio of sensitivities, but does not correspond to a change in basis matrix such as \textbf{G} to create a link between $\textbf{M}(\phir)$ and $\textbf{M}(\dphir)$.

A recent study by \citet{cisse2025} demonstrated that the interaction matrix measured around the phase residuals, \textbf{M}$(\dphir)$, does not permit the accurate retrieval of $\dphi$. The authors introduced a specific matrix that enables the correct reconstruction. To first order, the optical gains associated with this matrix can be expressed as
\begin{align}
\textbf{G}_\textbf{s}(\dphir) & = \frac{1}{2}\left( \mathbbm{1} + \dfrac{\text{diag}\left(\textbf{M}^\textsc{t}(\phir)\,\textbf{M}(\dphir)\right)}{\text{diag}\left(\vphantom{\phi}\textbf{M}^\textsc{t}(\phir)\,\textbf{M}(\phir)\right)}\right) \notag \\
                                        & = \frac{1}{2}\left(\vphantom{a^2} \mathbbm{1} + \textbf{G}(\dphir)\right)
,\end{align}
and are based on the quantity measured in the Eq.~\eqref{eq:OG_matricial_hyp}. Eventually it is remarkable that, in the three definitions proposed, \textbf{G}, $\textbf{G}_\textbf{K}$, and $\textbf{G}_\textbf{s}$, optical gains are understood and measured as a ratio of \textbf{M}$(\dphir)$ and \textbf{M}$(\phir)$, which in practice needs to be estimated. 

\subsection{Measuring the optical gains}
\label{sec:state-of-the-art-og}

The previous definitions of optical gains are only based on the knowledge of a ratio of interaction matrices. Other approaches have also been considered in order to estimate the optical gains using the telemetry of the closed-loop.

The CLOSE method, introduced by \citet{deo2019}, uses closed-loop telemetry to estimate an optimal modal loop gain that ensures system operation near the edge of stability, thereby maximising performance. This approach does not provide direct access to the optical gains values, but rather yields a combined estimate of the optimal modal loop gain. Its principal advantage lies in the fact that it requires no additional hardware and does not interfere with the science path. However, it has not yet been demonstrated on-sky.

An alternative, the optical gain telemetry loop (OGTL), proposed by \citet{esposito2015}, employs probe signals modulated on the deformable mirror during the closed loop. By comparing the reference amplitude with the amplitude recovered from telemetry data, the method enables measurement of optical gains. This technique, which also avoids the need for dedicated hardware, has been successfully tested and validated on-sky \citep{agapito2021}. Nevertheless, it is invasive, as the probes impact the scientific instruments, and it does not support frame-by-frame estimation, the update frequency being of the order of the hertz. Moreover, optical gains are measured only for a limited number of modes, with the remainder extrapolated.

A third approach, introduced by \citet{chambouleyron2020}, is the gain scheduling camera (GSC) method. It leverages focal-plane images in combination with the convolutional model described in \citet{fauvarque2019} to infer the optical gains. This technique requires additional hardware and a dedicated optical design to access the focal-plane images. Its main advantage is that it does not affect the science path and utilises only a small fraction of the light directed to the WFS, while enabling estimation of optical gains across the entire modal basis, potentially at a frame-by-frame cadence. The implementation of this concept is described in greater detail in Sect.~\ref{sec:gsc}.

\begin{table}[!ht]
	\caption{Comparison of state-of-the art methods for optical gain estimation.}
	\label{tab:og_state-of-the-art}
	\centering
	{\renewcommand{\arraystretch}{1.3}	
    \begin{tabular}{lccc}
		Method    & CLOSE & OGTL & GSC \\ \hline
		No loss of photons   & \checksymb	& \checksymb	& \unchecksymb \\
		Non-invasive         & \checksymb	& \unchecksymb	& \checksymb   \\
		OG only        		 & \unchecksymb	& \checksymb	& \checksymb   \\
		OG for all modes	 & —           	& \unchecksymb	& \checksymb   \\
		Real time     		 & \checksymb 	& \unchecksymb	& \checksymb   \\
		No additional optics & \checksymb 	& \checksymb	& \unchecksymb \\ \hline
	\end{tabular}}
\end{table}

A summary of the various approaches is presented in Tab.~\ref{tab:og_state-of-the-art}. The choice between strategies is inherently system-dependent, and involves a trade-off based on the error budget and the desired performance. In the context of extreme adaptive optics (XAO), and considering the PWFS operated away from its zero set point, the GSC appears to be a promising candidate. It provides estimates of the optical gains for the full modal basis at a high temporal rate, at the expense of diverting a small fraction of flux from the WFS path.

\section{The focal-plane assisted pyramid wavefront sensor on PAPYRUS}
\label{sec:FPAPWS}

\subsection{Principle of the GSC}
\label{sec:gsc}

\begin{figure}[!ht]
    \centering
    \includegraphics{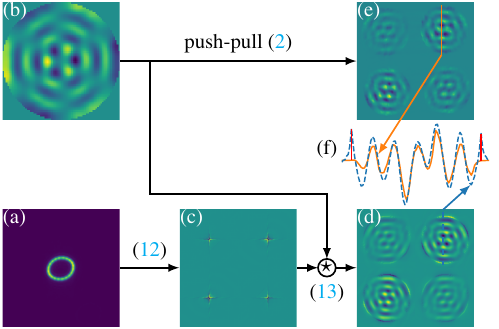}
    \caption{(a) Experimental focal-plane camera frame; (b) KL mode 80; (c) impulse response, in symmetrical logarithmic scale for visualisation purpose; (d) reconstructed PFWS response using the convolutional model; (e) experimental linear response, obtained through push-pull process; (f) 1D profile of one sub-pupil of experimental and simulated PWFS signal, with red lines indicating the edge effects due to the convolutional model.}
    \label{fig:model_convo}
\end{figure}

As was mentioned at the beginning of Sect.~\ref{sec:def_og}, the loss of sensitivity leading to the optical gains arises from the spreading of energy at the tip of the pyramid. Consequently, sensing this distribution using a focal-plane camera appears to be particularly valuable. By directing part of the flux towards a focal plane camera, it becomes possible to measure the intensity of the electromagnetic field conjugated to the one in the pyramid plane. Coupling the information acquired by the focal-plane camera, noted $\Omega(\dphir)$ (Fig.~\ref{fig:model_convo}a), with the Fourier-filtering analyser formalism introduced by \citet{fauvarque2019}, enables the extraction of an impulse response:
\begin{equation}
    \text{IR}(\dphir) = 2 \Im \left( \overline{\widehat{m}} \left(\widehat{m\ \Omega(\dphir)}\right) \right)
    \label{eq:ir}
,\end{equation}
with $\Im(\cdot)$ the imaginary part, $\overline{\cdot}$ the conjugate operator, $\widehat{\cdot}$ the Fourier transform, and $m$ the pyramid complex mask.

Assuming the linearity of the PWFS, which holds in the small phase regime, and translation invariance, the sensor can be modelled as a convolutional system. In practice, the translation invariance is not strictly satisfied but the hypothesis of a sliding pupil \citep{fauvarque2019} allows one to get closer to it and simplify the mathematical formalism. Under this framework, convolving the impulse response (Fig.~\ref{fig:model_convo}c) with a phase mode (Fig.~\ref{fig:model_convo}b) gives an approximation (Fig.~\ref{fig:model_convo}d) of the linear pyramid signal (Fig.~\ref{fig:model_convo}e):
\begin{equation}
    \delta I(\epsilon Z_i{;}\dphir) \approx \text{IR}(\dphir) \star Z_i
    \label{eq:delta_i_modele_convo}
,\end{equation}
with $\star$ the convolution product. The resulting signals shown in Fig.~\ref{fig:model_convo}c and Fig.~\ref{fig:model_convo}d are quite similar, but differ at the edges due to the approximations of the convolutional model described previously. This effect is clearly visible in Fig.~\ref{fig:model_convo}d and f, where the signal amplitude spikes at the limits of the pupil support. All the results shown in Fig.~\ref{fig:model_convo} were obtained experimentally using the PAPYRUS test bed, presented in Sect~\ref{sec:papyrus}.

The interaction matrix being the concatenation of the $\delta I$, the convolutional model enables one to estimate $\textbf{M}(\dphir)$, the missing piece for the estimation of the optical gains. In this work, we focus on the definition of the optical gains given in Eq.~(\ref{eq:OG_matricial_hyp}), which can be written \citep{chambouleyron2021} as
\begin{equation}
	\textbf{G}_{i}(\dphir) = \frac{\pscal{\text{IR}(\dphir) \star Z_i}{\text{IR}(\phir) \star Z_i}}{\pscal{\text{IR}(\phir) \star Z_i}{\text{IR}(\phir) \star Z_i}}
    \label{eq:og_gsc}
,\end{equation}
with $\pscal{\!\!\cdot}{\cdot\!\!}$ the inner product associated with the two-norm.

\begin{figure}[!ht]
    \centering
    \includegraphics{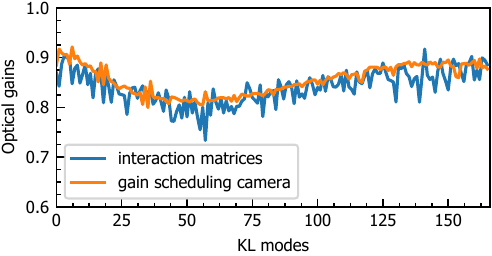}
    \caption{Experimental measurement of the optical gains using PAPYRUS internal source. The phase offset applied on the deformable mirror shown has an amplitude of 105~nm~RMS.}
    \label{fig:og_E2EvsGSC}
\end{figure}

As Eq.~(\ref{eq:og_gsc}) computes the optical gain for a specific mode, it must be applied across the entire modal basis. It is important to note that the artefacts -- such as the edge effects -- have only a negligible impact on the optical gains measurement. In fact, by taking the ratio of objects originating from the same model errors are, in this case, mitigated. As evidence of this statement, an experimental result in laboratory is presented in Fig.~\ref{fig:og_E2EvsGSC}, comparing the optical gains estimated via the push-pull interaction matrix measurements Eq.~(\ref{eq:OG_matricial_hyp}) and those obtained using the GSC. For this purpose, both methods are calibrated around a flat wavefront, resulting, respectively, in an interaction matrix and an impulse response. Afterwards, this calibration is repeated for the PWFS, using an extracted phase screen from AO telemetry of previous closed-loop as an offset applied on PAPYRUS deformable mirror -- allowing one to measure the optical gains considered as ground truth. The impulse response around these residuals is also measured delivering the optical gains using the GSC. The relative error measured between the optical gains of both methods, defined as
\begin{equation}
	e = \frac{\sqrt{\frac{1}{m}\sum_{i=1}^{m}(\textbf{G}_{\text{reference,}i} - \textbf{G}_{\text{measured},i})^2}}{\frac{1}{m}\sum_{i=1}^{m}\textbf{G}_{\text{reference},i}}
,\end{equation}
quantifies the error on optical gain estimates and is around 3.2\% in this situation. While no formal requirement has yet been established regarding the necessary precision of optical gain measurements, we aim for the most accurate estimation, since the goal is to perform the XAO operation. The GSC method, despite its underlying assumptions, is therefore suitable for this task as is demonstrated in \citet{chambouleyron2021} in a simulation that has now also been confirmed experimentally in the controlled environment of the laboratory. The optical gains curves follow the shape expected for a modulated PWFS \citep{deo2018a, chambouleyron2020}, with values decreasing until the spatial frequency of the mode matches the modulation radius and then increasing again. This is due to the dual behaviour of the modulated pyramid in sensitivity described in \citet{verinaud2004}. The choices made for the integration of the current GSC are presented in the following section alongside PAPYRUS test bed design.

\subsection{PAPYRUS: A development platform for AO concepts}
\label{sec:papyrus}

\begin{figure}[!ht]
	\centering
	\includegraphics{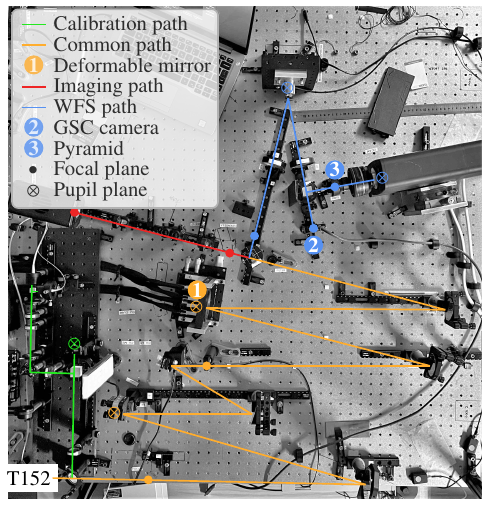}
	\caption{PAPYRUS AO test bed.}
	\label{fig:papyrus_bench}
\end{figure}

PAPYRUS \citep{fetick2023} is a pyramid-based AO system installed at the coudé focus of a 1.52~metre telescope at OHP. Its primary purpose is to serve as a test bed for on-sky AO concepts, components, and methods -- paving the way for technologies that will be implemented in VLT and ELT instruments. Its main characteristics are listed in Tab.~\ref{tab:papyrus} and its optical design is shown in Fig.~\ref{fig:papyrus_bench}.

As part of the current research and development activity, a GSC has been integrated on PAPYRUS to validate the concept in real conditions. It allows for direct comparisons to be made between on-sky results and realistic simulation outputs generated using a numerical twin of the test bed, thereby enabling a full validation of the method. This validation step is essential for operating the PWFS at a non-zero set point, as is required for applications such as those described in Sect.~\ref{sec:off-zero_cases}.

\begin{table}[!ht]
    \caption{Main characteristics of the AO test bed PAPYRUS.}
    \label{tab:papyrus}
    \centering
    \begin{tabular}{ll}
         Telescope &  \\
         \hspace{.25cm} Diameter & 1.52 m \\
         \hline Deformable Mirror & \\ 
         \hspace{.25cm} Number of actuators & 241 (cartesian 17x17)\\
         \hline Wavefront Sensing path & \\
         \hspace{.25cm} Spectral range & 400 -- 950 nm \\
         \hspace{.25cm} Modulation & 3 -- 10 $\lambda/D$ (usually 5) \\
         \hspace{.25cm} Pyramid field of view & 30'' \\
         \hspace{.25cm} GSC camera sampling & 2.7 pix/($\lambda/D$) \\
         \hspace{.25cm} GSC camera maximum field of view & 47.7''~x~35.8'' \\
         \hline Real Time Computer & Matlab homemade\\
         \hspace{.25cm} Frequency (w.o. OG monitoring) & up to 700~Hz \\
         \hspace{.25cm} Frequency (w. OG monitoring) & up to 400~Hz \\
         \hspace{.25cm} Delay & $\sim$3 frames @ 500~Hz \\
         \hline
    \end{tabular}
\end{table}

\begin{figure}[!ht]
    \centering
    \includegraphics{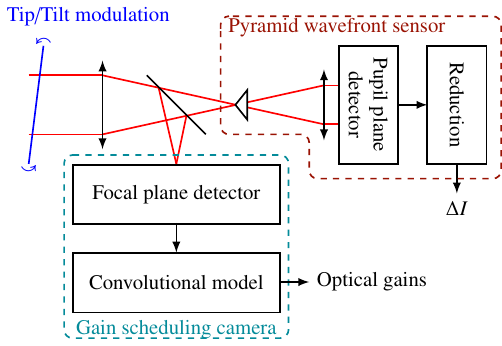}
    \caption{Schematic of the focal-plane assisted PWFS: combination of a PWFS branch and a GSC branch providing, respectively, the differential intensities and an estimate of optical gains.}
    \label{fig:fppwfs}
\end{figure}

The focal-plane assisted PWFS is the junction between the classical PWFS and the gain scheduling camera, whose principle diagram is presented in Fig.~\ref{fig:fppwfs}.

\subsection{Hardware implementation}

The focal-plane camera currently installed on PAPYRUS is a low-cost Thorlabs Zelux CS165CU. Its small dimensions facilitate straightforward integration into the optical bench. It is positioned downstream of one of the two outputs of the 50:50 beamsplitter cube. As is demonstrated in \citet{chambouleyron2021}, the estimation of optical gains remains accurate even for a sampling of the core of the PSF below the Shannon criterion, as long as the modulation radius still satisfies this criterion, and thus a sampling of at least 2 pixels of the modulation radius. However, the incident beam on the pyramid features a focal ratio of 14.5, and since no additional relay optics are used for the focal-plane camera, it inherits the same focal ratio. This results in a PSF sampling of 2.7~pix/($\lambda/D$). The only adjustable parameters -- without altering the optical design -- is the region of interest (ROI), which defines the field of view (FoV) of the camera, and the hardware binning. 

\subsubsection{Choice of the field of view and binning.}

\begin{figure}[!ht]
	\centering
	\includegraphics{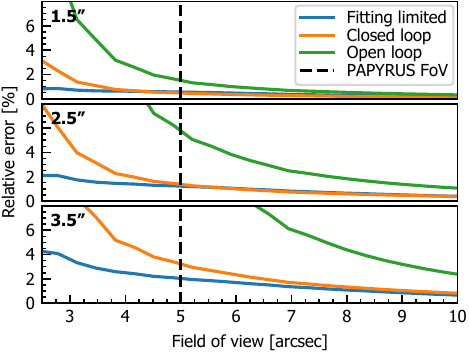}
	\caption{Simulated relative error, $e$, on the OG measurement with respect to the FoV with a reference FoV corresponding to the one of the PWFS (15~arcsec), for seeings of 1.5~arcsec (top), 2.5~arcsec (middle), and 3.5~arcsec (bottom). The dashed black line represents the chosen FoV on PAPYRUS.}
	\label{fig:rmse_fov}
\end{figure}

The natural approach when choosing the FoV of the focal plane camera is to match the one of the PWFS, ensuring that it sees the same perturbations. However, by its optical design the PAPYRUS PWFS has a FoV of 30~arcsec, over-sized for single-conjugate AO, where the seeing and anisoplanatism angle are of a few arcseconds. A study in simulation is conducted to estimate the minimal FoV required on the GSC to ensure accurate measurement of optical gains. The results, shown in Fig.~\ref{fig:rmse_fov}, are presented for various seeing conditions and across different regimes of correction. The middle graph corresponds to the median seeing typically observed at OHP. Based on this simulation, a FoV of 5~arcsec (58~$\lambda/D$ at 635~nm) ensures a relative error below 2\% in closed loop regime. This FoV is therefore kept for the focal plane camera, but can be adjusted in case of stronger or smaller seeing.

The current sampling, without binning, leads to images with a ROI of 160 by 160 pixels, which is suitable for fast computation. Indeed, the ROI defines the size of the matrices involved in the calculation of optical gains with Eq.~\eqref{eq:og_gsc} whose algorithmic complexity is discussed in section \ref{sec:complexity}. While reducing its size may increase computation time, it decrease the optical gains measurement accuracy, requiring a trade-off between the two.

In this spirit, hardware binning allows to reduce the ROI size without impacting the FoV. However, one should be careful to keep an ROI large enough to ensure the correct sampling of the influence functions, from which the modes $Z_i$ are computed.

\subsubsection{Impact of the detector quantisation.}

Due to its position, the focal-plane camera must offer a high dynamic range, as the PSF is highly contrasted -- especially in a closed-loop operation, in which most of the energy is concentrated in the core. To avoid saturation, acquisition parameters such as the exposure time and detector gain are adjusted accordingly. However, this can lead to part of the turbulent halo, spread over many pixels but with significantly lower intensity per pixel, falling below the camera's quantisation threshold. As a result, some energy is not recorded, introducing a bias into the global scaling and the shape of the measured optical gains.

\begin{figure}[!ht]
	\centering
	\includegraphics{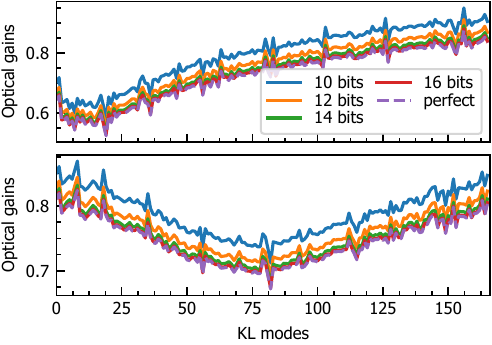}
	\caption{Simulated quantisation effect of the focal-plane detector on  optical gain estimation in closed-loop (seeing 2.5~arcsec) for modulation radii of 2~$\lambda$/D (top) and 5~$\lambda$/D (bottom).}
	\label{fig:quantization}
\end{figure}

This amplitude loss is estimated in the simulation by comparing the optical gains measured using different levels of quantisation with that of an ideal detector. The brightest pixel is filled at 80\% of its saturation. Results are shown in Fig.~\ref{fig:quantization}. For the 10-bit detector used in the PAPYRUS focal-plane camera, quantisation effects lead to a relative error of 5.8\% in the optical gain estimation in closed-loop conditions for a modulation radius of 5~$\lambda/D$ and up to 9.8\% with a modulation radius of 2~$\lambda/D$. This effect is thus further amplified for smaller modulation radii, as the modulated core is concentrated over fewer pixels, thereby increasing the contrast with the turbulent halo and increasing the impact of quantisation. However, this estimate is based on a noiseless simulation. In practice, the presence of strong noise on PAPYRUS focal-plane camera mitigates the threshold effect, most likely acting as a form of dithering. A more comprehensive analysis is required to fully characterise and understand this phenomenon. Regardless, a higher-quality detector with at least 12~bits, ideally 14~bits or 16~bits, would render the quantisation threshold effect negligible.

\subsection{Software algorithm implementation}
\label{sec:complexity}

Out of the four convolutions of Eq.~\eqref{eq:og_gsc}, three can be computed during the calibration process as they only require the knowledge of the impulse response in calibration leaving only one convolution -- $\text{IR}(\dphir) \star Z_i$ -- to compute for each of the $m$ modes of the basis at each iteration of the loop. Thus its complexity is, without considering the one of the impulse response:
\begin{equation}
	\mathcal{C}\eqref{eq:og_gsc} = m\times\texttt{C}\left(n^2\right) + m\times\texttt{SP}\left(n^2\right) 
,\end{equation}
with \texttt{C} and \texttt{SP}, respectively, the convolution and scalar product complexities and $n$ the number of pixels of one edge of the detector.
Using Fourier transform and Plancherel's theorem, it is possible to optimise the complexity, with
\begin{align}
\textbf{G}_{i}(\dphir) &= \frac{\pscal{\widehat{\text{IR}}(\dphir) \textcolor{ForestGreen}{\widecheck{\text{IR}}(\phir)}}{\textcolor{ForestGreen}{\widehat{Z_i} \widecheck{\hphantom{\,}Z_i\hphantom{\,}}\!}}}{\textcolor{ForestGreen}{\pscal{\widehat{\text{IR}}(\phir) \widecheck{\text{IR}}(\phir)}{\widehat{Z_i} \widecheck{\hphantom{\,}Z_i\hphantom{\,}}\!}}} \notag \\
    &= \frac{\vcenter{\hbox{$\sum_s$}}\, \widehat{\text{IR}}(\dphir) \textcolor{ForestGreen}{\widecheck{\text{IR}}(\phir) \widehat{Z_i} \widecheck{\hphantom{\,}Z_i\hphantom{\,}}\!}}{ \textcolor{ForestGreen}{\vcenter{\hbox{$\sum_s$}}\, \widehat{\text{IR}}(\phir) \widecheck{\text{IR}}(\phir) \widehat{Z_i} \widecheck{\hphantom{\,}Z_i\hphantom{\,}}\!}}
	\label{eq:og_gsc_opt}
,\end{align}
where $\widecheck{\cdot}$ denotes the inverse Fourier transform, $s$ the support of the impulse response, and a complexity of
\begin{equation}
	\mathcal{C}\eqref{eq:og_gsc_opt} = \texttt{FT}\left(n^2\right) + m\times\texttt{SP}\left(n^2\right)
,\end{equation}
with \texttt{FT} the 2D Fourier transform complexity, usually $\mathcal{O}(n^2 \log (n))$. This formulation reduces greatly the complexity, with only one Fourier transform being required to compute – all terms in green being pre-calculated in calibration – instead of $m$ convolution products. The $m$ scalar products left can be efficiently computed in parallel using multi-threading, significantly reducing the computation time. On state-of-the-art real-time computer (RTC) hardware, the computation with $m=195$ and $n=160$ can be done within a millisecond.

\begin{figure}[!ht]
	\centering
	\includegraphics{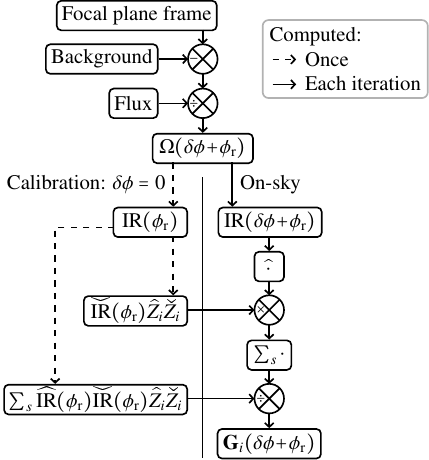}
	\caption{Optical gains processing pipeline, based on the implementation of Eq.~\eqref{eq:og_gsc_opt}.}
	\label{fig:pipeline_og}
\end{figure}

This calculation is achieved after the processing of the focal plane image. The background is removed and then the reduced image is normalised by its flux. From this processed image, the impulse response can be computed using Eq.~\eqref{eq:ir}, which is finally injected in Eq.~\eqref{eq:og_gsc_opt}. The different steps are summarised in Fig.~\ref{fig:pipeline_og}.

\section{On-sky validation}
\label{sec:onsky}
Since its first light in 2022 \citep{fetick2023}, PAPYRUS has undergone several changes. Initially operating in the visible for both the wavefront sensing and the imaging, it was later modified to switch the imaging arm in the infrared, hosting a focal plane imaging camera, a fibre-injection module and a port for a second stage Zernike-based AO system. The following on-sky results were obtained with these two configurations: observations of Vega were carried out entirely in the visible, while those of Altair and $\beta$~Peg were conducted with infrared imaging, the wavefront sensing remaining in the visible in both cases. The transition to an infrared imaging arm required the addition of numerous optical elements, which introduced NCPAs that were absent in the initial visible-only set-up and therefore need to be compensated for. The details of the on-sky acquisition considered in this paper are reported in Tab.~\ref{tab:stars}.

\begin{table}[!ht]
    \caption{Observed targets for optical gain measurement, and corresponding AO parameters.}
    \label{tab:stars}
	\centering
	\begin{tabular}{lccc}
		   & \textbf{Vega} & \textbf{Altair} & \textbf{$\beta$~Peg} \\ \hline
		Date (D/M/Y) & 24/05/23 & 26/07/24 & 26/07/24 \\ 
		CET (h:m) & 03:02 & 00:17 & 02:05 \\         
		Magnitude (V band) & 0.09 & 0.82 & 2.48 \\		
		Elevation & 76$^{\circ}$ & 52$^{\circ}$ & 57$^{\circ}$ \\		
		Zenithal seeing (500~nm) & 2.16" & 2.04" & 1.07" \\ \hline
		Frequency & 500 Hz & 400 Hz & 400 Hz \\
		Integrator gain & 0.5 & 0.4 & 0.3 \\
		Modulation & 5~$\lambda$/D & 5~$\lambda$/D & 5~$\lambda$/D \\
        Controlled KL modes & 192 & 166 & 166 \\
		Sensing wavelength & 635~nm & 635~nm & 635~nm \\
		Imaging wavelength & 635~nm & 1450~nm & 1450~nm \\
		GSC exposure time & 20~ms & 10~ms & 10~ms \\
		GSC binning & — & 2x2 & 2x2\\
		NCPAs compensation & 0 & $\lambda_\text{WFS}/4$  & $\lambda_\text{WFS}/4$ \\
		\hline
	\end{tabular}
\end{table}

\begin{figure*}[t]
    \centering
    \includegraphics[width=0.8\textwidth]{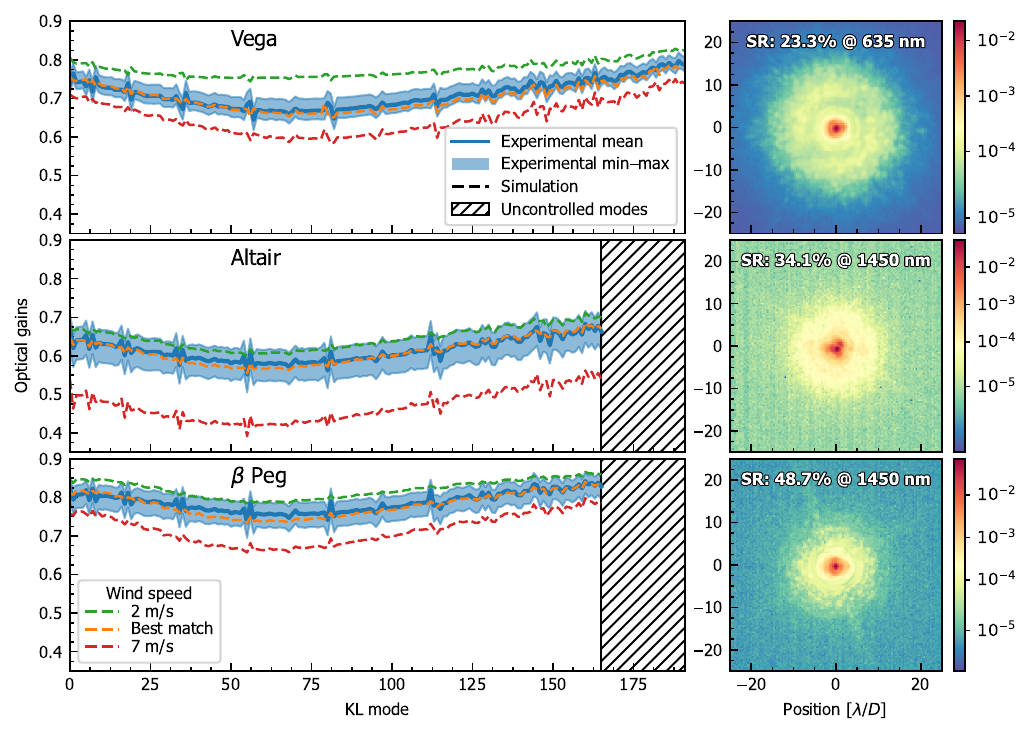}
    \caption{On-sky results of optical gains estimation during a closed-loop on Vega using PAPYRUS. Left: Optical gains. Experimental data are represented in solid blue, whereas dashed curves correspond to realistic simulations with different wind speeds. Right: Corresponding experimental long exposure PSF (50~seconds) on the imaging instrument.}
    \label{fig:og_all}
\end{figure*}

The GSC exposure time is either 20~ms or 10~ms, where the loop iteration duration is, respectively, 2~ms and 2.5~ms, showing that the optical gains are not estimated frame by frame. However, this limitation is purely technological as the focal plane camera is limited to a frame rate of 100~Hz and the current MATLAB RTC structure does not allow one to parallelise the phase reconstruction and the optical gains measurement.

For each observation we recorded the GSC telemetry, a long-exposure PSF, acquired on the imaging path, and all relevant AO-loop parameters. The long-exposure PSF, typically integrated over a few seconds, was fitted with the Psfao model from the Python library MAOPPY \citep{fetick2019a} to estimate the line-of-sight seeing at the imaging wavelength. This estimate was then converted to a zenithal seeing value at 500~nm using an elevation correction. The derived seeing values, together with the corresponding AO parameters, are reported in Tab.~\ref{tab:stars} and are used to run realistic simulations with a numerical twin of PAPYRUS using the Python library OOPAO \citep{heritier2023}. Based on the PAPYRUS modelling and the comparison with the experimental on-sky data, we assumed that the critical parameters are the seeing and the wind speed, since the temporal error dominates the AO error budget for a system operating at the frame rates given in Tab.~\ref{tab:stars}. Other contributors -- such as vibrations, dome seeing, the outer scale, $L_0$, and the $C_n^2$ profile -- were considered either negligible or effectively encapsulated in the PSF-derived seeing estimate.

The measured on-sky optical gains for the three observing cases are presented in Fig.~\ref{fig:og_all}. The right-hand column show the long-exposure PSF extracted from the imaging camera. These are the PSFs used to estimate the seeing, and the resulting seeing values are then used as an input to the simulations. The only remaining unknown parameter for each simulation is the wind speed integrated along the line of sight.

For each observing case, we carried out three simulations. Two simulations were computed for fixed wind speeds of 2 and 7~m.s$^{-1}$ (plotted in green and red, respectively) to provide reasonable lower and upper bounds that encompass the experimental measurements. A third simulation was made for a wind speed chosen individually for each observation (plotted in orange), manually adjusted so that the simulated optical gains curve best matches the measured curve (plotted in blue). The authors emphasise that the simulated data include limitations inherent to the model and are therefore used only as a sanity check rather than as proof of an absolute determination of the optical gains, because no ground truth is available.

\begin{figure}[!ht]
    \centering
    \includegraphics{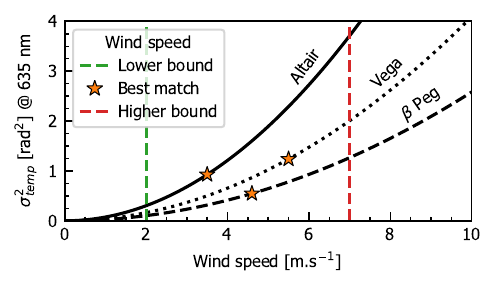}
    \caption{Theoretical temporal and fitting errors as a function of wind speed, computed using the AO parameters from Tab.~\ref{tab:stars} for each star. The colours match the ones of the simulated optical gains curves presented in Fig.~\ref{fig:og_all}.}
    \label{fig:temp_error_ws}
\end{figure}

The effect of wind speed on the theoretical temporal error is shown in Fig.~\ref{fig:temp_error_ws}, calculated using the parameters listed in Tab.~\ref{tab:stars}. An increase in the temporal error leads to a larger amplitude of the residual phase, and a larger residual phase amplitude enhances the non-linearity of the PWFS. Since optical gains are determined by the PWFS response, changes in the temporal error are reflected in changes of the measured optical gains. Consequently, a similar trend is expected, and is indeed observed, between the evolution of the optical-gain curves and that of the temporal error.

This relationship is particularly clear in the case of Altair. The temporal error increases by a factor of three when comparing the best matching wind speed to the upper-bound wind speed. This significant increase in the temporal error corresponds to a much larger increase in optical gains (i.e. values closer to zero) than those observed for the other two stars.

\begin{figure}[!ht]
    \centering
    \includegraphics{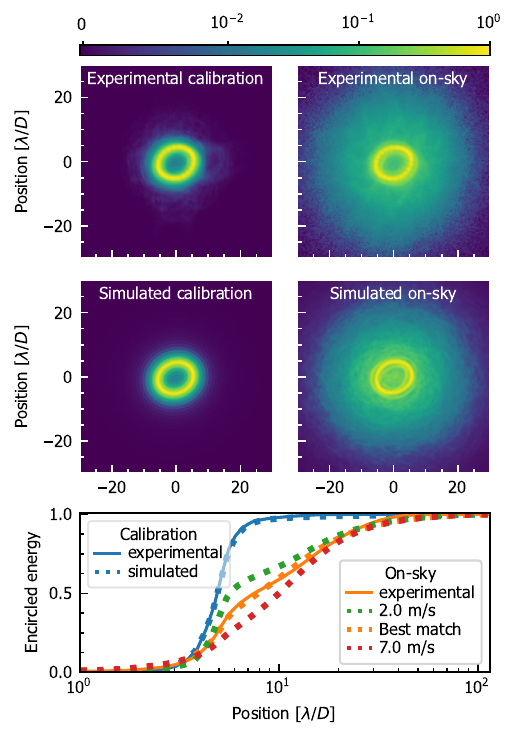}
    \caption{Experimental short-exposure (10~ms) focal-plane images (top row) and simulated focal-plane images (middle row) without noise and the wind speed of 5.5~m.s$^{-1}$. Bottom row: Encircled energies for experimental and simulated data for the different wind speed. Left: Calibration frame recorded on internal source. Right: On-sky frame recorded during closed-loop operation on Vega.}
    \label{fig:gsc_cal_and_sky_expe_synt}
\end{figure}

To further investigate the validity of the results, an additional analysis is provided in Fig.~\ref{fig:gsc_cal_and_sky_expe_synt}. The simulated and experimental GSC frames obtained for the vase of Vega are shown side by side, together with their associated encircled-energy profiles. Outside the AO correction radius, all optical gains curves overlap, which corroborates the seeing conditions used in the simulations being representative of those on-sky. Within the AO correction radius, and in particular at low spatial frequencies, the influence of wind speed becomes apparent: the encircled energies for the different wind speed are spread apart, and only the best matching value (dashed orange) reproduces the on-sky data closely.

Taken together, these elements indicate that the model used for the best-matching simulations is sufficiently representative of the observing conditions and that the derived estimates of the optical gains are meaningful. Moreover, the overall shape of the optical-gain curve is consistent with previously published results \citep{deo2018a,chambouleyron2020}.

\section{Conclusion}

In this work, we have demonstrated the feasibility of accurately measuring optical gains both in laboratory and on-sky using the GSC method, building on the theoretical and simulation works initiated in \citet{chambouleyron2020,chambouleyron2021}. This on-sky validation marks a significant step towards operational use of the focal-plane assisted PWFS, allowing the PWFS to work around a non-zero set point, with ongoing studies focused on stability and precision in such configurations. In particular, it will be applied to accurately compensate for NCPAs in order to maximise coupling efficiency into a single-mode fibre, feeding  the high-resolution VIPA fibre spectrograph \citep{carlotti2022}, recently coupled to PAPYRUS. It also opens the possibility of deploying this enhanced WFS as a first-stage sensor in future XAO systems, such as PCS. In this context, PAPYRUS is currently undergoing yet another upgrade, including a faster RTC based on shared memory \citep{cetre2023}, an improved focal-plane detector for the GSC, and a second-stage Zernike-based AO system.

While the present study focuses on the PWFS, the GSC remains applicable to any FFWFS, as it is derived from its analytical formalism, providing great versatility. Moreover, the GSC approach opens up the possibility of accessing more than only optical gains thanks to the focal-plane information that it brings. We can think of applications such as the monitoring of extended objects, assessing the AO correction quality, retrieving the PSF seen by the WFS, extracting information on low-order aberrations such as the low-wind effect, and even including the entire signal in the reconstruction. These features complement the off-zero operation of the PWFS and may significantly contribute to optimising AO performance.

   \begin{acknowledgements} 
This work benefited from the support of the French National Research Agency (ANR) with WOLF (ANR-18-CE31-0018), APPLY (ANR-19-CE31-0011) and LabEx FOCUS (ANR-11-LABX-0013); the Programme Investissement Avenir F-CELT (ANR-21-ESRE-0008), the Action Spécifique Haute Résolution Angulaire (ASHRA) of CNRS/INSU co-funded by CNES, the ECOS-CONYCIT France-Chile cooperation (C20E02), the ORP-H2020 Framework Programme of the European Commission’s (Grant number 101004719), STIC AmSud (21-STIC-09), the Région Sud and the french government under the France 2030 investment plan with the CASSIOPEE project, and as part of the Initiative d'Excellence d'Aix-Marseille Université -A*MIDEX, program number AMX-22-RE-AB-151.
   \end{acknowledgements}   
   
   \bibliographystyle{aa}
   \bibliography{AA55909-25}

\end{document}